\documentclass[preprint]{elsarticle} % options: preprint or review

\makeatletter
	\def\ps@pprintTitle{%
 	\let\@oddhead\@empty
	\let\@evenhead\@empty
	\def\@oddfoot{\centerline{\thepage}}%
	\let\@evenfoot\@oddfoot}
\makeatother

\usepackage{etoolbox}
\patchcmd{\MaketitleBox}{\footnotesize\itshape\elsaddress\par\vskip36pt}{\footnotesize\itshape\elsaddress\par\parbox[b][36pt]{\linewidth}{\vfill\hfill\textnormal{\today}\hfill\null\vfill}}{}{}%
\patchcmd{\pprintMaketitle}{\footnotesize\itshape\elsaddress\par\vskip36pt}{\footnotesize\itshape\elsaddress\par\parbox[b][36pt]{\linewidth}{\vfill\hfill\textnormal{\today}\hfill\null\vfill}}{}{}%

\usepackage[colorlinks=true,%
            breaklinks=true,%
            linkcolor=blue,
            urlcolor=blue,%
            citecolor=blue,%
            pdftitle={Title of paper}, 
            pdfkeywords={Keywords},	
            pdfauthor={Authors},		
            bookmarksopen=false,
            pdfpagemode=UseNone]{hyperref}
\usepackage[margin = 0.9in]{geometry}
\usepackage[T1]{fontenc}
\usepackage[english]{babel}
\usepackage[latin1]{inputenc} % Can cause compiling issues
\usepackage{mathtools}
\usepackage{siunitx}
\usepackage{amsmath}
\usepackage{amsfonts}
\usepackage{amsthm}
\usepackage{amssymb}
\usepackage{array}
\usepackage{algorithm} %ctan.org\pkg\algorithms
\usepackage{algorithmicx}
\usepackage{algpseudocode}
\usepackage{subcaption}
\usepackage{siunitx}
\usepackage{amsmath}
\usepackage{mathrsfs}
\usepackage{xpatch}

\usepackage{color}
\usepackage{url}
\usepackage{cleveref}
\usepackage{graphicx}
\usepackage{breakcites}
\usepackage{soul} % for using the \ul command with linebreak
\usepackage{enumitem}
\usepackage[title,titletoc,toc]{appendix}

\newtheoremstyle{mytheoremstyle}{5pt}{5pt}{\itshape}{}{\bfseries}{.}{.5em}{} 
\theoremstyle{mytheoremstyle}

\newtheoremstyle{myremarkstyle}{3pt}{3pt}{\itshape}{}{\bfseries}{.}{.5em}{} 
\theoremstyle{myremarkstyle}

\usepackage{pifont}% http://ctan.org/pkg/pifont
\newcommand{\cmark}{\ding{51}}%
\newcommand{\xmark}{\ding{55}}%

\newtheorem{proposition}{Proposition}

\newtheorem{definition}{Definition}

{\noindent {\textbf{Proof}:} }%
{\hfill $\Box$ \\ }

\newcommand{\bit}{\begin{itemize}}
\newcommand{\eit}{\end{itemize}}
\newcommand{\ben}{\begin{enumerate}}
\newcommand{\een}{\end{enumerate}}

\newcommand{\Hquad}{\hspace{0.5em}} 
 
\begin{document}
\begin{frontmatter}
    \title{Conservative Closures of the Vlasov-Poisson Equations Based on Symmetrically Weighted Hermite Spectral Expansion}
    \author[ucsd,lanl]{Opal Issan\corref{cor1}}\ead{oissan@ucsd.edu}
    \author[lanl]{Oleksandr Koshkarov}
    \author[ga]{Federico D. Halpern}
    \author[ucsd]{Boris Kramer}
    \author[lanl]{Gian Luca Delzanno}
    
    \cortext[cor1]{Corresponding author}
    \address[ucsd]{Department of Mechanical and Aerospace Engineering, University of California San Diego, La Jolla, CA, USA}
    \address[lanl]{T-5 Applied Mathematics and Plasma Physics Group, Los Alamos National Laboratory, Los Alamos, NM, USA}
    \address[ga]{General Atomics, P.O. Box 85608, San Diego, CA, USA}
    \begin{abstract}
        We derive conservative closures for the Vlasov-Poisson equations discretized in velocity via the symmetrically weighted Hermite spectral expansion. We demonstrate that no closure can simultaneously restore the conservation of mass, momentum, and energy in this formulation. The properties of the analytically derived conservative closures of each conserved quantity are validated numerically by simulating an electrostatic benchmark problem: the Langmuir wave.  Both the numerical results and analytical analysis indicate that closure by truncation (i.e. setting the last Hermite moment to zero) is the most suitable conservative closure for the symmetrically weighted Hermite formulation. 
    \end{abstract}
\end{frontmatter}
%%%%%%%%%%%%%%%%%%%%%%%%%%%%%%%%%%%%%%%%

%%%%%%%%%%%%%%%%%%%%%%%%%%%%%%%%%%%%%%%
\section{Introduction} 
%%%%%%%%%%%%%%%%%%%%%%%%%%%%%%%%%%%%%%%
% What is the kinetic moment closure problem?
The kinetic Vlasov equation describes the evolution of the particle distribution function in seven dimensions: three spatial, three velocity, and time coordinates. The reduced fluid description is derived by taking moments of the kinetic equation with respect to the velocity coordinates, which evolve the density, fluid momentum, pressure, etc. This process reduces the number of independent variables to space and time, removing the velocity dependence, which renders the fluid description more computationally tractable. However, the fluid evolution equation for the $n$th moment involves the ($n+1$)th moment, resulting in an \textit{infinite} moment hierarchy. 

% What are the most common ways to come up with closures?
To obtain a finite hierarchy, one needs to define a moment \textit{closure}. This is commonly done by truncating the moment hierarchy and relating the last moment to the lower-order moments, or by assuming that the last moment is zero (known as \textit{closure by truncation}), all while trying to accurately approximate the interplay between the microscopic kinetic and macroscopic fluid processes.
For strongly collisional plasma, common closures assume that the heat flux is proportional to the gradient of the temperature~\cite{spitzer_1953_collisional, braginskii_1965_collisional}, however, such four-moment closures are valid for small Knudsen number ($\text{Kn} \ll 1$) plasma flows, where the particle mean-free-path is much smaller than the system's characteristic length scale. Traditional collisionless ($\text{Kn}\sim1$) closures typically aim to mimic specific kinetic phenomena, e.g. Landau damping~\cite{hammet_1990_closure, hammet_1992_closure, wang_2019_landau_closure, hunana_2019_landau_closure} or magnetic reconnection~\cite{ng_2020_reonnection_closure}. Additionally, recent studies use neural networks~\cite{maulik_2020_nn, ma_2020_nn, bois_2022_nn, christlieb_2024_hyperbolic_nn, han_2019_nn}, sparse regression~\cite{alves_2022_sg, donaghy_2023_sg}, extrapolation~\cite{gibelli_2006_closure, joyce_1971_closure}, and system-theoretic model reduction~\cite{gillot_2021_bt} to derive collisionless and weakly collisional kinetic closures.

% What about Hermite discretization and its connection to closure
Most spectral methods for kinetic equations have focused on using Hermite functions in velocity space, weighted by a Gaussian exponential function~\cite{kormann_2021_sw, holloway_1996_sw, schumer_1998_sw_aw, koshkarov_2021_sps, delzanno_2015_sps, camporeale_sps_2016}. The Hermite spectral expansion is particularly effective because it can approximate near-Maxwellian distributions with relatively few basis functions. There are mainly two classes of Hermite discretization. The first, introduced by Grad in 1949~\cite[\S 4]{grad_1949_sw}, evolves the Hermite basis dynamically in space and time by its dependence on fluid velocity and temperature. In this method, conservation of mass, momentum, and energy is enforced directly by solving the fluid equations, independently of the closure. The closure only impacts hyperbolicity, which must be maintained to avoid unphysical dissipation in the strictly advective collisionless system~\cite{cai_2013_hyperbolic, struchtrup_2003_regularization, koellermeier_2014_hyperbolic}. The second approach uses a fixed Hermite basis through either an \textit{asymmetrically weighted} (AW) or \textit{symmetrically weighted} (SW) expansion. The AW expansion conserves mass, momentum, and energy regardless of the closure term due to its fluid-kinetic coupling, where the first few Hermite coefficients describe the fluid moments~\cite{schumer_1998_sw_aw}. Its main drawback is that it suffers from numerical instability. Conversely, the SW expansion is numerically stable but lacks conservation properties and fluid-kinetic coupling, such that all Hermite coefficients are required to describe the fluid quantities~\cite{schumer_1998_sw_aw, holloway_1996_sw, issan_2023_antisymmetric}. Therefore, the closure of the SW expansion impacts its conservation properties. 

% What is this short note/paper about?
In this work, we investigate alternative conservative closures for the one-dimensional collisionless Vlasov-Poisson equations discretized in velocity via the SW Hermite spectral expansion~\cite{holloway_1996_sw, issan_2023_antisymmetric}. We derive new closures that restore the SW conservation of total mass, momentum, or energy. Additionally, we examine the proposed conservative closures preservation of hyperbolicity, and anti-symmetry (important for stability purposes) of the Vlasov equation. We verify numerically the analytically derived conservative closures on a classic benchmark problem: the Langmuir wave. 

% What is the outline of this short note/paper?
This paper is organized as follows. First, section~\ref{sec:vlasov-poisson-SW} briefly reviews the one-dimensional collisionless Vlasov-Poisson equations and their velocity discretization via the SW Hermite spectral expansion. Section~\ref{sec:closure-conservative} presents three conservative closures: (1) closure by truncation, which we previously rigorously analyzed in~\cite{issan_2023_antisymmetric},  (2) closure by momentum, and (3) closure by energy. Section~\ref{sec:structure-preservation} discusses the closures' ability to preserve the Vlasov equation hyperbolicity and anti-symmetric structure. We test numerically the analytic derivations of the closure properties on the Langmuir wave test in section~\ref{sec:numerical-results}. Lastly, in section~\ref{sec:conclusions}, we discuss conclusions and further considerations.

%%%%%%%%%%%%%%%%%%%%%%%%%%%%%%%%%%%%%%%
\section{Vlasov-Poisson Equations: Symmetrically Weighted Hermite Spectral Expansion in Velocity}\label{sec:vlasov-poisson-SW}
%%%%%%%%%%%%%%%%%%%%%%%%%%%%%%%%%%%%%%%
% describe the governing equations and boundary conditions we consider
We investigate the one-dimensional Vlasov-Poisson equations, which describe the interaction between collisionless charged particles coupled to a self-consistent electric field.
The plasma is characterized by the non-negative and bounded particle distribution function $f^{s}(x, v, t)$ in phase space. This function represents the number of particles in the phase space element $\mathrm{d}x \mathrm{d}v$ at a specific time $t\in \Omega_{t} \subseteq \mathbb{R}_{\geq 0}$. Here, the plasma species are identified by the label $s$.
We consider a periodic spatial coordinate $x$ within the domain $\Omega_{x} = [0, \ell]$, where $\ell$ denotes the length of the spatial domain. The time domain is $\Omega_{t} = [0, t_{f}]$, with $t_{f} \in \mathbb{R}_{\geq 0}$ indicating the final time. Additionally, the velocity coordinate is unbounded, such that $\Omega_{v} = \mathbb{R}$.
The one-dimensional (normalized) Vlasov-Poisson equations are then expressed as follows:
\begin{alignat}{3}
    \left(\frac{\partial}{\partial t} + v \frac{\partial}{\partial x}+ \frac{q^{s}}{m^{s}} E(x, t) \frac{\partial}{\partial v}\right) f^{s}(x, v, t)  &= 0,  \qquad &&\text{in } \Omega_{x} \times \Omega_{v} \times \Omega_{t},\Hquad \forall s, \label{vlasov-continuum}\\
    \frac{\partial E(x, t)}{\partial x} &= \sum_{s} q^{s} \int_{\mathbb{R}} f^{s}(x,v, t) \mathrm{d}v, \qquad &&\text{in } \Omega_{x} \times \Omega_{t}, \label{poisson-continuum}
\end{alignat}
where $E(x, t)$ is the electric field, $m^{s}$ is the mass of species $s$, and $q^{s}$ is the charge of species $s$.  All quantities in the Vlasov-Poisson equations~\eqref{vlasov-continuum}--\eqref{poisson-continuum} are normalized in the centimeter-gram-seconds system of units as follows:
\begin{equation*}
    \qquad q^{s} \coloneqq \frac{q_{d}^{s}}{e}, \qquad m^{s} \coloneqq \frac{m_{d}^{s}}{m^{e}}, \qquad t \coloneqq t_{d} \omega^{e}, \qquad x \coloneqq \frac{x_{d}}{\lambda^{e}}, \qquad f^{s} \coloneqq f_{d}^{s}\frac{v^{e}}{n^{e}},  \qquad E \coloneqq E_{d} \frac{e \lambda^{e}}{T^{e}}, \qquad v \coloneqq \frac{v_{d}}{v^{e}},
\end{equation*}
where the subscript `$d$' indicates the dimensional quantities, $e$ is the positive elementary charge, $m^{e}$ is the electron mass,  $\omega^{e} \coloneqq \sqrt{4 \pi e^2 n^{e}/m^{e}}$ is the electron plasma frequency, $n^{e}$ is the reference electron density, $\lambda^{e} \coloneqq \sqrt{T^{e}/4\pi e^{2}n^{e}}$ is the electron Debye length, $T^{e}$ is a reference electron temperature,  $v^{e} \coloneqq \sqrt{T^{e}/m^{e}}$ is the electron thermal velocity. 
To ensure the uniqueness of the solution, we enforce the condition that $\int_{\Omega_{x}}E(x, t) \mathrm{d} x = 0$ for all $t \in \Omega_{t}$.

We employ a truncated spectral expansion in velocity based on the SW Hermite basis to approximate the particle distribution function, i.e.
\begin{equation}\label{spectral-ansatz}
    f^{s}(x, v,t) \approx f^{s, N_{v}}(x, v, t) = \sum_{n=0}^{N_{v}-1} C_{n}^{s}(x, t) \psi_{n}(\xi^{s}), \qquad \xi^{s}(v) \coloneqq \frac{v-u^{s}}{\alpha^{s}}.
\end{equation}
The velocity coordinate $v$ is shifted by $u^{s}\in \mathbb{R}$ and scaled by $\alpha^{s} \in \mathbb{R}_{>0}$~\cite{tao_1993_hermite}, which are two tunable parameters specified by the user. 
The Hermite parameters $u^{s}$ and $\alpha^{s}$ correspond to the characteristic mean flow and thermal velocity for each species $s$, as the zeroth-order SW Hermite basis functions $\psi_0(\xi_s)$ resemble a Maxwellian distribution. These parameters are typically set based on the initial condition's characteristic mean flow and thermal velocity.
The SW Hermite basis function of degree $n \in \mathbb{N}_{0}$ is
\begin{equation}\label{hermite-basis-function}
    \psi_{n}(\xi^{s}) \coloneqq (\sqrt{\pi} 2^n n!)^{-\frac{1}{2}} \mathcal{H}_{n}(\xi^{s}) \exp{\left(-\frac{(\xi^{s})^2}{2}\right)},
\end{equation}
and $\mathcal{H}_{n}(\xi^{s})$ is the ``\textit{physicist}'' Hermite polynomial~\cite[\S 22]{abramowitz_1964_math} of degree $n \in \mathbb{N}_{0}$. The SW Hermite basis functions form an orthogonal basis, i.e. $\int_{\mathbb{R}} \psi_{n}(\xi^{s}) \psi_{m}(\xi^{s}) \mathrm{d} \xi^{s} = \delta_{n,m}$, where $\delta_{n,m}$ is the Kronecker delta function. After Galerkin projection and exploiting the orthogonality and recurrence relation of the SW Hermite basis functions, see~\cite[\S 2.2]{issan_2023_antisymmetric} for more details, we derive a system of partial differential equations for the expansion coefficients, such that
\begin{align}\label{pde-dynamics}
    \frac{\partial C_{n}^{s}(x, t)}{\partial t} &+ \underbrace{\frac{\partial }{\partial x}\left(\alpha^{s} \sqrt{\frac{n}{{2}}} C_{n-1}^{s}(x, t) + \alpha^{s} \sqrt{\frac{n+1}{2}} C_{n+1}^{s}(x, t) + u^{s}  C_{n}^{s}(x, t)\right)}_{\text{advection } v\partial_{x} f^{s}}\\
    &+\underbrace{\frac{q^{s}}{m^{s}\alpha^{s}} 
 E(x,t)\left(-\sqrt{\frac{n}{2}}C_{n-1}^{s}(x, t) + \sqrt{\frac{n+1}{2}} C_{n+1}^{s}(x, t)\right)}_{{\text{acceleration } q^{s}/m^{s} E\partial_{v} f^{s}}}= 0, \nonumber
\end{align}
with the convention that $C_{n<0}^{s}(x, t) = 0$ and a closure, yet to be specified, of the following form
\begin{equation}\label{closure-arbitrary}
C_{N_{v}}^{s}(x, t) = \mathcal{F}\left(C_{N_{v}-1}^{s}(x, t), E(x, t)\right).
\end{equation}
Note that the SW Hermite closure introduces an error in both the advection $(v \partial_{x} f^{s})$ and acceleration $(q^{s}/m^{s} E \partial_{v} f^{s})$ terms, as highlighted in Eq.~\eqref{pde-dynamics}.
The Poisson equation~\eqref{poisson-continuum} after Galerkin projection is given by 
\begin{equation}\label{poisson-pde}
    \frac{\partial E(x, t)}{\partial x} = \sum_{s} q^{s} \alpha^{s} \sum_{n=0}^{N_{v}-1} \mathcal{I}^{0}_{n} C_{n}^{s}(x, t)\qquad \mathrm{with} \qquad \mathcal{I}^{0}_{n} = \int_{\mathbb{R}} \psi_{n}(\xi^{s}) \mathrm{d} \xi^{s}.
\end{equation}
The integral $\mathcal{I}^{0}_{n} \in \mathbb{R}$ recursive properties are as follows
\begin{equation}\label{I0-recursive}
    \mathcal{I}^{0}_{0} = \sqrt{2} \pi^{\frac{1}{4}}, \qquad \mathcal{I}_{1}^{0} = 0, \qquad \mathrm{and} \qquad \mathcal{I}_{n}^{0} = \sqrt{\frac{n-1}{n}} \mathcal{I}_{n-2}^{0} \qquad \mathrm{for} \qquad n \geq 2. 
\end{equation}

%%%%%%%%%%%%%%%%%%%%%%%%%%%%%%%%%%%%%%%
\section{Conservative Closures for the Symmetrically Weighted Hermite Formulation} \label{sec:closure-conservative}
%%%%%%%%%%%%%%%%%%%%%%%%%%%%%%%%%%%%%%%
The main question we seek to answer is: \textit{Is there a closure in the form of Eq.~\eqref{closure-arbitrary} that improves the mass, momentum, energy, or Casimir (e.g., $\mathcal{L}_{2}$ norm) conservation properties of the SW Hermite formulation?} We begin by deriving the mass, momentum, and energy fluid equations and their respective drift rates with an arbitrary closure, see~\cite[\S 4.1]{issan_2023_antisymmetric} for a more detailed derivation for the closure by truncation.
Then, we set the drift rates to zero and derive a conservative closure.  The proposed closures are (1) closure by truncation, (2) closure by momentum, and (3) closure by energy. The closures result in a system of partial differential-algebraic equations for the coefficients motivated by the fluid equations, which are the first three moments with respect to the velocity monomials $,1, v, v^2,$ of the Vlasov equation~\eqref{vlasov-continuum}. 

%%%%%%%%%%%%%%%%%%%%%%%%%%%%%%%%%%%%%%%
\subsection{Mass Conservative Closure (Equivalent to Closure by Truncation)} \label{sec:mass-conservation}
%%%%%%%%%%%%%%%%%%%%%%%%%%%%%%%%%%%%%%%
The mass of species $s$ is the zeroth moment of the particle distribution function
\begin{equation}\label{mass-definition}
    M_{0}^{s}(x, t) \coloneqq m^{s}  \int_{\mathbb{R}}f^{s, N_{v}}(x, v, t) \mathrm{d} v  = m^{s} \alpha^{s} \sum_{n=0}^{N_{v}-1} \mathcal{I}^{0}_{n} C_{n}^{s}(x, t),
\end{equation}
where the integral $\mathcal{I}^{0}_{n} \in \mathbb{R}$ is defined in Eq.~\eqref{poisson-pde}.
Then, by taking the time partial derivative of Eq.~\eqref{mass-definition} and inserting Eq.~\eqref{pde-dynamics}, the SW continuity equation is given by 
\begin{equation}\label{continuity-SW}
    \frac{\partial M_{0}^{s}(x, t)}{\partial t} + \frac{\partial M_{1}^{s}(x, t)}{\partial x} = -q^s E(x, t) 
\begin{cases}
\sqrt{\frac{N_{v}}{2}} \mathcal{I}^{0}_{N_v-1} C_{N_v}^{s}(x, t) \qquad &\text{if } N_{v} \text{ is odd},\\
\sqrt{\frac{N_{v}-1}{2}} \mathcal{I}^{0}_{N_{v}-2} C^{s}_{N_v-1}(x, t) \qquad &\text{if } N_{v} \text{ is even},\\
\end{cases} 
\end{equation}
where the right-hand side introduces a closure error stemming from the finite SW Hermite discretization in velocity. The momentum of species $s$ is the first moment of the particle distribution function 
\begin{equation}\label{momentum-definition}
    M_{1}^{s}(x, t) \coloneqq m^{s} \int_{\mathbb{R}} v f^{s, N_{v}}(x, v, t) \mathrm{d} v  = \alpha^{s} m^{s} \sum_{n=0}^{N_{v}-1} \mathcal{I}^{1}_{n} C_{n}^{s}(x, t), \qquad \text{with} \qquad \mathcal{I}^{1}_{n} \coloneqq \int_{\mathbb{R}} v \psi_{n}(\xi^{s}) \mathrm{d}\xi^{s}.
\end{equation}
The integral $\mathcal{I}_{n}^{1} \in \mathbb{R}$ is related to the integral $\mathcal{I}_{n}^{0} \in \mathbb{R}$ in Eq.~\eqref{I0-recursive} by 
\begin{equation*}
    \mathcal{I}_{n}^{1} = \begin{cases}
    \alpha^{s}\left(\sqrt{\frac{n+1}{2}} \mathcal{I}^{0}_{n+1} + \sqrt{\frac{n}{2}} \mathcal{I}^{0}_{n-1} \right) = \alpha^{s}\sqrt{2n} \mathcal{I}^{0}_{n-1} = \alpha^{s} \sqrt{2(n+1)} \mathcal{I}_{n+1}^{0} \qquad &\text{if } n \geq 1\text{ and odd,}\\
    u^{s}\mathcal{I}^{0}_{n} \qquad &\text{if } n\geq 0 \text{ and even.}
    \end{cases}
\end{equation*}
Therefore, by Eq.~\eqref{continuity-SW}, the continuity equation is only satisfied via the \textit{closure by truncation}, i.e. $C_{N_{v}}^{s}(x, t) = 0$, and odd $N_{v}$. By integrating Eq.~\eqref{continuity-SW} with respect to space, the total mass drift rate of species $s$ is given by 
\begin{alignat}{3}\label{change-in-mass-sw-closure}
\frac{\mathrm{d}}{\mathrm{d}t} \int_{0}^{\ell} M_{0}^{s}(x, t) \mathrm{d} x= - q^s
\begin{cases}
\sqrt{\frac{N_{v}}{2}} \mathcal{I}^{0}_{N_v-1} \int_{0}^{\ell} E(x, t) C_{N_v}^{s}(x, t) \mathrm{d} x\qquad &\text{if } N_{v} \text{ is odd},\\
\sqrt{\frac{N_{v}-1}{2}} \mathcal{I}^{0}_{N_{v}-2} \int_{0}^{\ell} E(x, t) C^{s}_{N_v-1}(x, t) \mathrm{d} x\qquad &\text{if } N_{v} \text{ is even}.\\
\end{cases}
\end{alignat}
The closure cannot impact the mass drift rate if $N_{v}$ is even. However, if $N_{v}$ is odd, then setting $\int_{0}^{\ell}E(x, t) C_{N_v}^{s}(x, t)\mathrm{d} x = 0$ causes the total mass drift rate to vanish.  The closure by truncation, i.e. $C_{N_{v}}^{s}(x, t) = 0$, is one example of a closure resulting in a total mass conservative formulation and the \textit{only} closure that satisfies the continuity equation. 

%%%%%%%%%%%%%%%%%%%%%%%%%%%%%%%%%%%%%%%
\subsection{Momentum Conservative Closure} \label{sec:momentum-conservation}
%%%%%%%%%%%%%%%%%%%%%%%%%%%%%%%%%%%%%%%
We proceed to derive the SW Hermite fluid momentum equation by taking the time partial derivative of Eq.~\eqref{momentum-definition} and inserting Eq.~\eqref{pde-dynamics}, such that
\begin{align}\label{momentum-equation}
    \frac{\partial M^{s}_{1}(x, t)}{\partial t} + \frac{\partial M^{s}_{2}(x, t)}{\partial x} &- \frac{q^{s}}{m^{s}} M_{0}^{s}(x, t) E(x, t) = \\
    &-q^{s} E(x, t) \begin{cases}
\mathcal{I}^{0}_{N_{v}-1}  \left(u^{s} \sqrt{\frac{N_{v}}{2}}  C_{N_{v}}^{s}(x, t)+ \alpha^{s} N_v C_{N_v-1}^{s}(x, t) \right) \qquad &\text{if } N_{v} \text{ is odd},\\
\mathcal{I}^{0}_{N_v-2} \sqrt{\frac{N_v-1}{2}} \left(u^{s}  C_{N_v-1}^{s}(x, t) +\alpha^{s}  \sqrt{2N_{v}}C_{N_{v}}^{s}(x, t)\right) \qquad&\text{if } N_{v} \text{ is even}, 
\end{cases}\nonumber
\end{align}
where the right-hand side introduces a closure error stemming from the finite SW Hermite expansion in velocity. Mass $M^{s}_{0}(x, t)$ is defined in Eq.~\eqref{mass-definition}, momentum $M^{s}_{1}(x, t)$ is defined in Eq.~\eqref{momentum-definition}, and kinetic energy $\frac{1}{2}M_{2}(x, t)$ is defined by the second-moment of the particle distribution function
\begin{equation}\label{kinetic-energy-definition}
    M_{2}^{s}(x, t) \coloneqq m^{s}\int_{\mathbb{R}} v^{2} f^{s, N_{v}}(x, v, t) \mathrm{d} v = m^{s} \alpha^{s}   \sum_{n=0}^{N_{v}-1} \mathcal{I}_{n}^{2}  C_{n}^{s}(x, t), \qquad \text{with} \qquad \mathcal{I}_{n}^{2} \coloneqq \int_{\mathbb{R}} v^{2} \psi_{n}(\xi^{s}) \mathrm{d} \xi^{s}.
\end{equation}
The integral $\mathcal{I}_{n}^{2} \in \mathbb{R}$ is related to the integral $\mathcal{I}_{n}^{0} \in \mathbb{R}$ in Eq.~\eqref{I0-recursive} by 
\begin{equation*}
    \mathcal{I}_{n}^{2} = \begin{cases}
        2\alpha^s u^s \left( \sqrt{\frac{n+1}{2}} \mathcal{I}^{0}_{n+1} + \sqrt{\frac{n}{2}} \mathcal{I}^{0}_{n-1}\right) = 2u^{s} \mathcal{I}_{n}^{1} \qquad &\text{if } n\geq 1 \text{ and odd,}\\
        (\alpha^{s})^2 \left(\sqrt{\frac{(n+1)(n+2)}{4}} \mathcal{I}^{0}_{n+2} + \left( \frac{2n+1}{2} + \left(\frac{u^s}{\alpha^s}\right)^2\right) \mathcal{I}^{0}_{n} + \sqrt{\frac{n(n-1)}{4}} \mathcal{I}^{0}_{n-2}\right)\qquad &\text{if }n\geq 0 \text{ and even.}
    \end{cases}
\end{equation*}
Then, the total momentum drift rate is given by integrating Eq.~\eqref{momentum-equation} in space for all species $s$, i.e. 
\begin{equation}\label{change-in-momentum-sw-closure}
    \frac{\mathrm{d}}{\mathrm{d} t} \sum_{s} \int_{0}^{\ell} M_{1}^{s}(x, t) \mathrm{d} x = -\sum_{s} q^{s} \begin{cases}
\mathcal{I}^{0}_{N_{v}-1} \int_{0}^{\ell} E(x, t) \left(u^{s} \sqrt{\frac{N_{v}}{2}}  C_{N_{v}}^{s}(x, t)+ \alpha^{s} N_v C_{N_v-1}^{s}(x, t) \right) \mathrm{d} x \qquad &\text{if } N_{v} \text{ is odd},\\
\mathcal{I}^{0}_{N_v-2} \sqrt{\frac{N_v-1}{2}} \int_{0}^{\ell} E(x, t) \left(u^{s}  C_{N_v-1}^{s}(x, t) +\alpha^{s}  \sqrt{2N_{v}}C_{N_{v}}^{s}(x, t)\right) \mathrm{d} x\qquad&\text{if } N_{v} \text{ is even}.\\
\end{cases}
\end{equation}
Therefore, the fluid momentum equation is satisfied and the momentum drift rate vanishes if
\begin{equation}\label{momentum-closure}
    C_{N_v}^{s}(x, t) = -C_{N_v-1}^{s}(x, t) \begin{cases}
\frac{\alpha^s\sqrt{2N_{v}}}{u^{s}} \qquad &\text{if } N_{v} \text{ is odd},\\
\frac{u^s}{\alpha^{s} \sqrt{2N_{v}}} \qquad&\text{if } N_{v} \text{ is even}.\\
\end{cases}
\end{equation}
We refer to this closure as \textit{closure by momentum}. For both cases of $N_{v}$ being odd or even, the closure term $C_{N_{v}}^{s}(x, t)$ is proportional to the last Hermite expansion coefficient $C_{N_{v}-1}^{s}(x, t)$. It is also important to mention that when $N_{v}$ is odd this closure requires the velocity shifting parameter $u^{s} \neq 0$ for all species $s$, which can be easily imposed by performing a Galilean transformation. In the case of $N_{v}$ being even and $u^{s}=0$, the closure by truncation also satisfies the momentum equation and thus preserves the total momentum. 

%%%%%%%%%%%%%%%%%%%%%%%%%%%%%%%%%%%%%%%
\subsection{Energy Conservative Closure} \label{sec:energy-conservation}
%%%%%%%%%%%%%%%%%%%%%%%%%%%%%%%%%%%%%%%
Next, we derive the energy fluid equation by taking the time partial derivative of Eq.~\eqref{kinetic-energy-definition} and inserting Eq.~\eqref{pde-dynamics}, i.e.
\begin{align}
   \frac{1}{2} \frac{\partial M_{2}^{s}(x, t)}{\partial t} + &\frac{1}{2} \frac{\partial M_{3}^{s}(x, t)}{\partial x} - \frac{q^{s}}{m^{s}} M_{1}^{s}(x, t) E(x, t) =  \label{energy-equation}\\
   &-q^{s}  E(x, t) \begin{cases}
\mathcal{I}^{0}_{N_v-1}  \left(u^{s} \alpha^{s} N_{v}
    C^{s}_{N_v-1}(x, t) + \sqrt{\frac{N_v}{2}} \left(\eta^{s}_{N_{v}} C_{N_{v}}^{s}(x, t) \right) \right) \qquad &\text{if } N_{v} \text{ is odd},\nonumber\\
\mathcal{I}^{0}_{N_v-2} \left( \sqrt{\frac{N_v-1}{2}} \left(\gamma^{s}_{N_{v}} C^{s}_{N_v-1}(x, t) \right)  + u^{s} \alpha^{s} \sqrt{N_{v}(N_{v}-1)}C_{N_{v}}^{s}(x, t)\right)
&\text{if } N_{v} \text{ is even,}\nonumber
\end{cases}
\end{align}
where $\eta^{s}_{N_{v}} = \frac{(2N_v +1) (\alpha^{s})^2 + (u^s)^2}{2}$ and $\gamma^{s}_{N_{v}} = \frac{(2N_v -1) (\alpha^{s})^2 + (u^s)^2}{2}$.
Therefore, the right-hand side introduces closure error originating from the SW Hermite velocity discretization. The energy flux $\frac{1}{2}M^{s}_{3}(x, t)$ is the third moment of the particle distribution function, such that 
\begin{equation*}
    M^{s}_{3}(x, t) \coloneqq m^{s}\int_{\mathbb{R}} v^3 f^{s, N_{v}}(x, v, t) \mathrm{d} v = m^{s} \alpha^{s} \sum_{n=0}^{N_{v}-1} \mathcal{I}_{n}^{3} C_{n}^{s}(x, t), \qquad \text{with} \qquad 
    \mathcal{I}_{n}^{3} \coloneqq \int_{\mathbb{R}} v^{3} \psi_{n}(\xi^{s}) \mathrm{d} \xi^{s}.
\end{equation*}
The total energy is the sum of kinetic and potential energies, such that $\mathcal{E}(t) \coloneqq \frac{1}{2} \sum_{s} \int_{0}^{\ell} M_{2}^{s}(t) \mathrm{d}x + \frac{1}{2} \int_{0}^{\ell} E(x, t)^{2} \mathrm{d} x$. From Eqns.~\eqref{pde-dynamics} and~\eqref{poisson-pde}, the total energy drift rate is
\begin{equation}\label{change-in-total-energy-sw-closure}
\frac{\mathrm{d}\mathcal{E}}{\mathrm{d}t} = \sum_{s} q^{s}  \int_{0}^{\ell} E(x, t) \begin{cases}
\mathcal{I}^{0}_{N_v-1}  \left(u^{s} \alpha^{s} N_{v}
    C^{s}_{N_v-1}(x, t) + \sqrt{\frac{N_v}{2}} \left(\eta^{s}_{N_{v}} C_{N_{v}}^{s}(x, t) + \mu_{N_{v}}^{s}(x, t)\right) \right) \mathrm{d} x\qquad &\text{if } N_{v} \text{ is odd},\\
\mathcal{I}^{0}_{N_v-2} \left( \sqrt{\frac{N_v-1}{2}} \left(\gamma^{s}_{N_{v}} C^{s}_{N_v-1}(x, t) +  \mu^{s}_{N_{v}-1}(x, t)\right)  + u^{s} \alpha^{s} \sqrt{N_{v}(N_{v}-1)}C_{N_{v}}^{s}(x, t)\right) \mathrm{d} x
&\text{if } N_{v} \text{ is even,}
\end{cases}
\end{equation}
where $\mu^{s}_{n}(x, t) =  \frac{q^{s}}{m^{s}}\int_{0}^{x} E(y, t) C_{n}^{s}(y, t) \mathrm{d} y$. The extra term involving $\mu_{n}^{s}(x,t)$ stems from integrating the SW Poisson equation~\eqref{poisson-pde} to arrive at the SW Ampere's equation. 
The total energy drift rate in Eq.~\eqref{change-in-total-energy-sw-closure} is zero, if
\begin{alignat}{3}\label{energy-closure}
    C_{N_{v}}^{s}(x, t) = \begin{cases}\frac{-1}{\eta_{N_{v}}^{s}}\left( u^{s}\alpha^{s}\sqrt{2N_{v}}C_{N_{v}-1}^{s}(x, t) +\mu_{N_{v}}^{s}(x, t)\right) \qquad&\text{if } N_{v} \text{ is odd}, \\
\frac{-1}{u^{s}\alpha^{s} \sqrt{2N_{v}}} \left(\gamma^{s}_{N_{v}}C^{s}_{N_v-1}(x, t) + \mu_{N_{v}-1}^{s}(x, t) \right)  \qquad&\text{if } N_{v} \text{ is even}. 
\end{cases}
\end{alignat}
We refer to this closure as \textit{closure by energy}. When $N_{v}$ is odd, the energy-conserving closure term in Eq.~\eqref{energy-closure} satisfies a linear Volterra integral equation of the second kind~\cite[\S 16.1]{arfken_2013_book}, which can be solved numerically (after discretization in space), for example using the trapezoidal rule. Similar to the momentum closure, the energy closure with even $N_{v}$ requires the velocity shifting parameter $u^{s} \neq 0$ for all species $s$, which again can be imposed by performing a Galilean transformation. In the case of $N_{v}$ being odd and $u^{s}=0$, closure by truncation also satisfies the energy equation and conserves the total energy. 

Note that closure by energy in Eq.~\eqref{energy-closure} does not satisfy the energy equation, i.e. Eq.~\eqref{energy-equation} with the vanishing right-hand-side, but conserves the total energy (the sum of kinetic and potential energies). This arises due to the inequivalence between the Poisson and Ampere equations after SW Hermite discretization in velocity, which requires the continuity equation to be satisfied. Recall that the continuity equation is only satisfied for closure by truncation and odd $N_{v}$. Alternatively, by solving the Vlasov-Ampere system via the SW Hermite expansion in velocity, the closure by energy will conserve the total energy and satisfy the energy equation, yet then closure by momentum will not be able to conserve the total momentum and satisfy the momentum equation simultaneously.  

%%%%%%%%%%%%%%%%%%%%%%%%%%%%%%%%%%%%%%%
\subsection{$\mathcal{L}_{2}$ Norm Conservative Closure}\label{sec:L2-conservation}
%%%%%%%%%%%%%%%%%%%%%%%%%%%%%%%%%%%%%%%
The $\mathcal{L}^{s}_{2}$ norm of the particle distribution function of species $s$ is also an invariant of the Vlasov-Poisson equations~\eqref{vlasov-continuum}-\eqref{poisson-continuum}, which is defined as 
\begin{equation*}
    \mathcal{L}_{2}^{s}(t) \coloneqq \int_{0}^{\ell} \int_{\mathbb{R}} f^{s, N_{v}}(x, v, t)^2 \mathrm{d} v \mathrm{d} x = \alpha^{s} \sum_{n=0}^{N_{v}-1} C^{s}_{n}(x, t)^2.
\end{equation*}
We denote the total species norm as $\mathcal{L}_{2}(t) = \sum_{s} \mathcal{L}_{2}^{s}(t)$. Following Eq.~\eqref{pde-dynamics}, the $\mathcal{L}^{s}_{2}$ norm drift rate is 
\begin{equation*}
    \frac{\mathrm{d} \mathcal{L}_{2}^{s}}{\mathrm{d} t} = - \sqrt{2N_{v}} \frac{q^{s}}{m^{s}} \int_{0}^{\ell} C_{N_{v}-1}^{s}(x, t)E(x, t) C_{N_v}^{s}(x, t)\mathrm{d} x.
\end{equation*}
Thus, the closure by truncation, $C_{N_{v}}^{s}(x, t) = 0$, results in a $\mathcal{L}^{s}_{2}$ norm conserving closure. 

%%%%%%%%%%%%%%%%%%%%%%%%%%%%%%%%%%%%%%%
\section{Closure Structure Preservation: Hyperbolicity and Anti-symmetry}\label{sec:structure-preservation}
%%%%%%%%%%%%%%%%%%%%%%%%%%%%%%%%%%%%%%%
We check below if the three closures preserve hyperbolicity and anti-symmetry. We summarize the conservation and structure-preserving properties of the three closures in Table~\ref{tab:closure-list}.
Let us begin by writing the evolution equations of the expansion coefficients in Eq.~\eqref{pde-dynamics} in vector form:
\begin{equation}\label{pde-vector-form}
    \frac{\partial \mathbf{C}^{s}(x, t)}{\partial t} + \mathbf{J}^{s} \frac{\partial \mathbf{C}^{s}(x, t)}{\partial x} = \mathbf{S}^{s}(x, t)\mathbf{C}^{s}(x, t), 
\end{equation}
where $\mathbf{C}^{s}(x, t) \coloneqq \left[C_{0}^{s}(x, t), C_{1}^{s}(x, t), \ldots, C_{N_{v}-1}^{s}(x, t)\right]^{\top} \in \mathbb{R}^{N_{v}}$. The flux Jacobian matrix $\mathbf{J}^{s} \in \mathbb{R}^{N_{v} \times N_{v}}$ with a general closure in the form of Eq.~\eqref{closure-arbitrary}, is a real symmetric tridiagonal matrix:
\begin{equation}\label{flux-jacobian-matrix}
    \mathbf{J}^{s} \coloneqq \begin{bmatrix} 
    u^{s}&  \alpha^{s} \sqrt{\frac{1}{2}}& 0 &  \ldots & 0 \\
    \alpha^{s} \sqrt{\frac{1}{2}} & u^{s} & \alpha^{s} \sqrt{\frac{2}{2}} &  \ldots & 0\\
    & &  & & \\
    & \ddots & \ddots &  \ddots&   \\
    & &  & & \\
    0 & 0 &  \alpha^{s} \sqrt{\frac{N_v-2}{2}} & u^{s} & \alpha^{s} \sqrt{\frac{N_v-1}{2}}\\
    0 & 0 & 0 & \alpha^{s} \sqrt{\frac{N_v-1}{2}}  & u^{s} + \sigma^{s}
\end{bmatrix} \in \mathbb{R}^{N_{v} \times N_{v}},
\end{equation}
where $\sigma^{s} \in \mathbb{R}$ linearly relates the closure term $C_{N_{v}}^{s}(x, t)$ to $C_{N_{v}-1}^{s}(x, t)$.
Similarly, by Eq.~\eqref{pde-dynamics}, the source operator with a general closure in the form of Eq.~\eqref{closure-arbitrary} is given by 
\begin{equation}\label{source-operator}
    \mathbf{S}^{s}(x, t) \coloneqq \frac{q^{s}}{m^{s} \alpha^{s}} E(x, t) \begin{bmatrix} 
    0&  -\sqrt{\frac{1}{2}}& 0 &  \ldots & 0 \\
    \sqrt{\frac{1}{2}} &0 & -\sqrt{\frac{2}{2}} &  \ldots & 0\\
    & &  & & \\
    & \ddots & \ddots &  \ddots&   \\
    & &  & & \\
    0 & 0 &  \sqrt{\frac{N_v-2}{2}} & 0 & -\sqrt{\frac{N_v-1}{2}}\\
    0 & 0 & 0 & \sqrt{\frac{N_v-1}{2}}  & \beta^{s}(x, t)
\end{bmatrix} \in \mathbb{R}^{N_{v} \times N_{v}},
\end{equation}
where $\beta^{s}(x, t)$ depends on the closure function~\eqref{closure-arbitrary}.

\begin{table}
\caption{A list of proposed SW closures and their respective conservation properties.}
\centering
\begin{tabular}{c |c |c | c }
\textbf{closure properties}& \shortstack{\textbf{closure by truncation} \\ $C_{N_{v}}^{s}(x, t) = 0$}  & \shortstack{\textbf{closure by momentum} \\ Eq.~\eqref{momentum-closure}} & \shortstack{\textbf{closure by energy} \\ Eq.~\eqref{energy-closure}}\\
\hline
mass conservation & if $N_{v}$ is odd  &  \xmark  & \xmark \\
\hline
momentum conservation & if $N_{v}$ is even and $u^{s} = 0, \Hquad \forall s$ & \cmark & \xmark\\
\hline
energy conservation & if $N_{v}$ is odd and $u^{s} = 0, \Hquad \forall s$& \xmark & \cmark\\
\hline 
$\mathcal{L}_{2}$ norm conservation & \cmark  & \xmark & \xmark \\
\hline
hyperbolicity & \cmark & \cmark & \cmark\\
\hline
anti-symmetry & \cmark  & \xmark & \xmark
\end{tabular}
\label{tab:closure-list}
\end{table}

%%%%%%%%%%%%%%%%%%%%%%%%%%%%%%%%%%%%%%%
\subsection{Hyperbolicity}
%%%%%%%%%%%%%%%%%%%%%%%%%%%%%%%%%%%%%%%
It is important for a moment/spectral closure to preserve the hyperbolic nature of the continuous Vlasov equation~\eqref{vlasov-continuum} to maintain the strictly advective (non-dissipative) nature of the equations. We start our discussion by recalling the definition of a hyperbolic system.
\begin{definition}[Hyperbolic System]\label{hyperbolic-defintion}
    If the flux Jacobian matrix $\mathbf{J}^{s} \in \mathbb{R}^{N_{v} \times N_{v}}$ in Eq.~\eqref{pde-vector-form} has only real eigenvalues and a complete set of eigenvectors, then the system is called hyperbolic. 
\end{definition}
\noindent
The following theorem asserts that all general closures in the form of Eq.~\eqref{closure-arbitrary} preserve the hyperbolicity of the Vlasov equation~\eqref{vlasov-continuum}. 
\begin{proposition}[{\cite[\S 7]{parlett_1998_symmetric}}]\label{eigenvalue-proposition}
    A real symmetric tridiagonal matrix has real eigenvalues and a complete set of eigenvectors if all sub-diagonal elements are nonzero.
\end{proposition}
\noindent
Therefore, from Eq.~\eqref{flux-jacobian-matrix}, the hyperbolicity is enforced via the SW Hermite spectral ansatz in Eq.~\eqref{spectral-ansatz} and the general closure relation in Eq.~\eqref{closure-arbitrary}. Note that allowing the closure term to depend on lower-order expansion coefficients will break the symmetry of the Jacobian flux matrix, and consequently can break hyperbolicity. 

So far we have treated the case with constant $\alpha^{s}$ and $u^{s}$. However, there is a strong mathematical and physical motivation to adapt spatially and temporally the Hermite shifting and scaling parameters, denoted as $u^{s}$ and $\alpha^{s}$ in Eq.~\eqref{hermite-basis-function}. The mathematical considerations stem from improving the convergence of the Hermite spectral discretization. The physical motivation stems from the Hermite basis function in Eq.~\eqref{hermite-basis-function} resembling a Maxwellian distribution (thermodynamic equilibrium), such that $u^{s}$ mimics the characteristic mean flow and $\alpha^{s}$ mimics the thermal velocity. Therefore, the Hermite parameters can be adapted in space and time based on the evolution of such macroscopic quantities. An adaptive approach based on evolving the macroscopic quantities was first proposed in the seminal paper by Grad~\cite{grad_1949_sw}, which suffers from a lack of hyperbolicity. On the other hand, recent work by~\cite{pagliantini_2023_adaptive} showed that the adaptivity in time of the AW Hermite parameters (i.e. $u^{s}$ and $\alpha^{s}$) does not alter the governing equations of the expansion coefficients and relies on projecting the expansion coefficients \textit{at a fixed instant in time} of the Hermite parameters. This allows the adaptive in-time AW and SW formulations to remain hyperbolic. The adaptivity in space of the Hermite parameters introduces new terms in the form of $C_{n}^{s}(x, t) \frac{\partial}{\partial x} \alpha^{s}(x)$ and $C_{n}^{s}(x, t) \frac{\partial}{\partial x} u^{s}(x)$ as additional ``source'' terms. These terms do not directly impacts the Jacobian flux matrix and similar to time-adaptivity require a Hermite projection step onto the new basis \textit{at a fixed instant in time}~\cite{koshkarov_2022_abstract}. Therefore, we anticipate that by following the proposed adaptive approach for the AW Hermite method in~\cite{pagliantini_2023_adaptive, koshkarov_2022_abstract}, a space and time adaptive approach of the SW Hermite method also preserves the system's hyperbolicity. 

%%%%%%%%%%%%%%%%%%%%%%%%%%%%%%%%%%%%%%%
\subsection{Anti-symmetry}
%%%%%%%%%%%%%%%%%%%%%%%%%%%%%%%%%%%%%%%
Anti-symmetry (also known as skew-symmetry) is an additional important structure for a spectral method to preserve. Aside from capturing the correct dynamics, an anti-symmetric discretization guarantees unconditional numerical stability~\cite{dai_2024_quadratic_AW, issan_2023_antisymmetric}. Our previous work~\cite[\S 3.2]{issan_2023_antisymmetric} explains in detail the anti-symmetric structure of the Vlasov advection operator. In short, it requires the flux Jacobian matrix in Eq.~\eqref{flux-jacobian-matrix} to be symmetric, i.e. $\mathbf{J}^{s} = (\mathbf{J}^{s})^{\top}$, and the source operator in Eq.~\eqref{source-operator} to be anti-symmetric, i.e. $\mathbf{S}^{s}(x, t) = -\mathbf{S}^{s}(x, t)^{\top}$. As previously stated in~\cite[\S 3.2.2]{issan_2023_antisymmetric}, anti-symmetry is only preserved via the closure by truncation, otherwise $\mathbf{S}^{s}(x, t) \neq -\mathbf{S}^{s}(x, t)^{\top}$. 
The violation of anti-symmetry stems from $\beta^{s}(x, t) \neq 0$ in Eq.~\eqref{source-operator}. 
Additionally, the adaptivity in time of the Hermite parameters following the work by~\cite{pagliantini_2023_adaptive} does not change the governing equations and will preserve anti-symmetry for the closure by truncation. However, adaptivity in space breaks anti-symmetry by introducing non-zero terms on the main diagonal of the source operator.

%%%%%%%%%%%%%%%%%%%%%%%%%%%%%%%%%%%%%%%
\section{Numerical Results} \label{sec:numerical-results}
%%%%%%%%%%%%%%%%%%%%%%%%%%%%%%%%%%%%%%%
We study numerically the closures listed in Table~\ref{tab:closure-list} on the Langmuir wave benchmark problem. We initialize the particle distribution functions as 
\begin{equation*}
    f^{e}(x, v, t=0) = \frac{1 + \epsilon \cos(x)}{\sqrt{2\pi}\alpha^{e}} \exp\left(-\frac{1}{2} \left( \frac{v - u^{e}}{\alpha^{e}}\right)^2\right) \qquad \mathrm{and} \qquad f^{i}(x, v, t=0) = \frac{1}{\sqrt{2\pi}\alpha^{i}} \exp\left(-\frac{1}{2} \left( \frac{v - u^{i}}{\alpha^{i}}\right)^2\right),
\end{equation*}
where $s=`e$' denotes electrons and $s=`i$' denotes ions, $\epsilon=0.01$ is the amplitude of the initial electron perturbation, $\alpha^{e} = 0.1$ and $\alpha^{i} = \sqrt{1/1836}$ is the average thermal velocity, $u^{e} = u^{i} = 1$ is the average fluid velocity,  and $\ell = 2\pi$ is the normalized length of the system. We set a realistic ion-to-electron mass ratio $m^{i}/m^{e}= 1836$. We employ the Fourier spectral discretization in space of Eq.~\eqref{pde-dynamics}, which we previously derived in~\cite[Appendix A]{issan_2023_antisymmetric}. We use the second-order implicit midpoint integrator with the unpreconditioned Jacobian-Free-Newton-Krylov method~\cite{knoll_jfnk_2004} with absolute error and relative error set to $10^{-12}$. The total number of Fourier spectral terms in space is $N_{x} = 101$ and the time step is $\Delta t = 0.1$. 

\begin{figure}
    \centering
     \begin{subfigure}[b]{0.49\textwidth}
         \centering
         \caption{$\mathcal{L}_{2}$ conservation}
         \includegraphics[width=\textwidth]{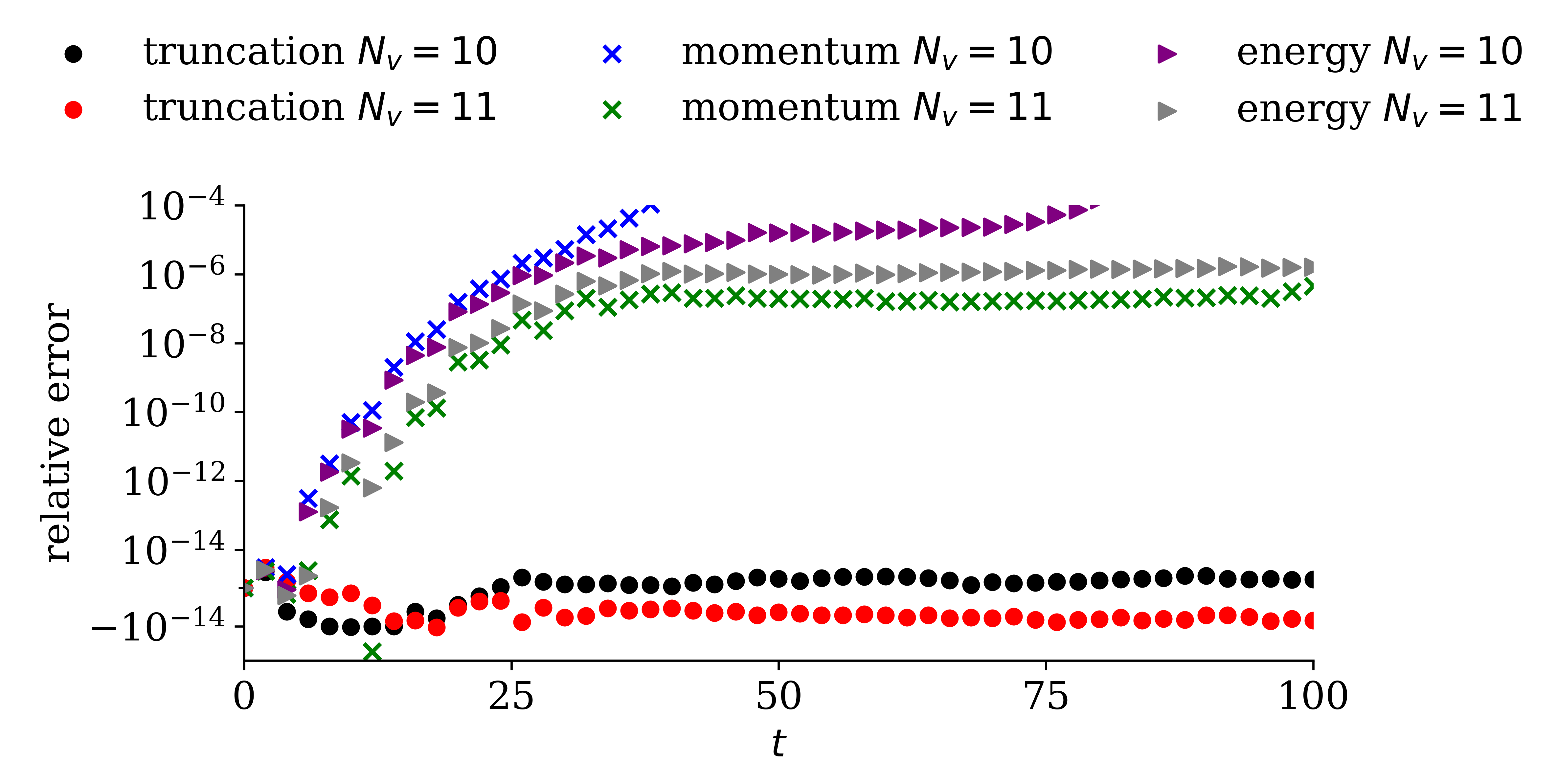}
         \label{fig:closure-by-truncation-10}
     \end{subfigure}
    \begin{subfigure}[b]{0.49\textwidth}
         \centering
         \caption{mass conservation}
         \includegraphics[width=\textwidth]{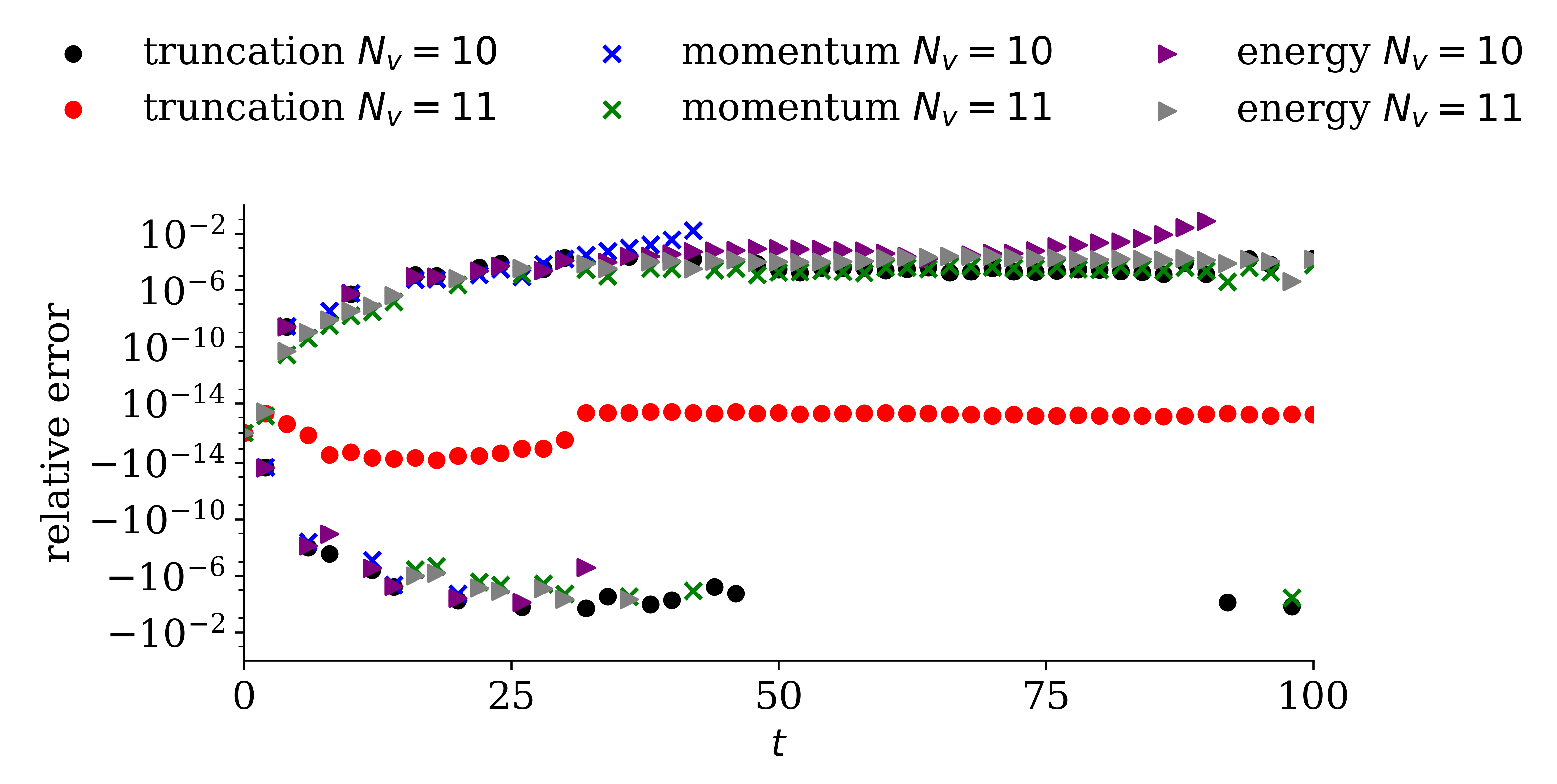}
         \label{fig:closure-by-truncation-11}
     \end{subfigure}
    \begin{subfigure}[b]{0.49\textwidth}
         \centering
         \caption{momentum conservation}
         \includegraphics[width=\textwidth]{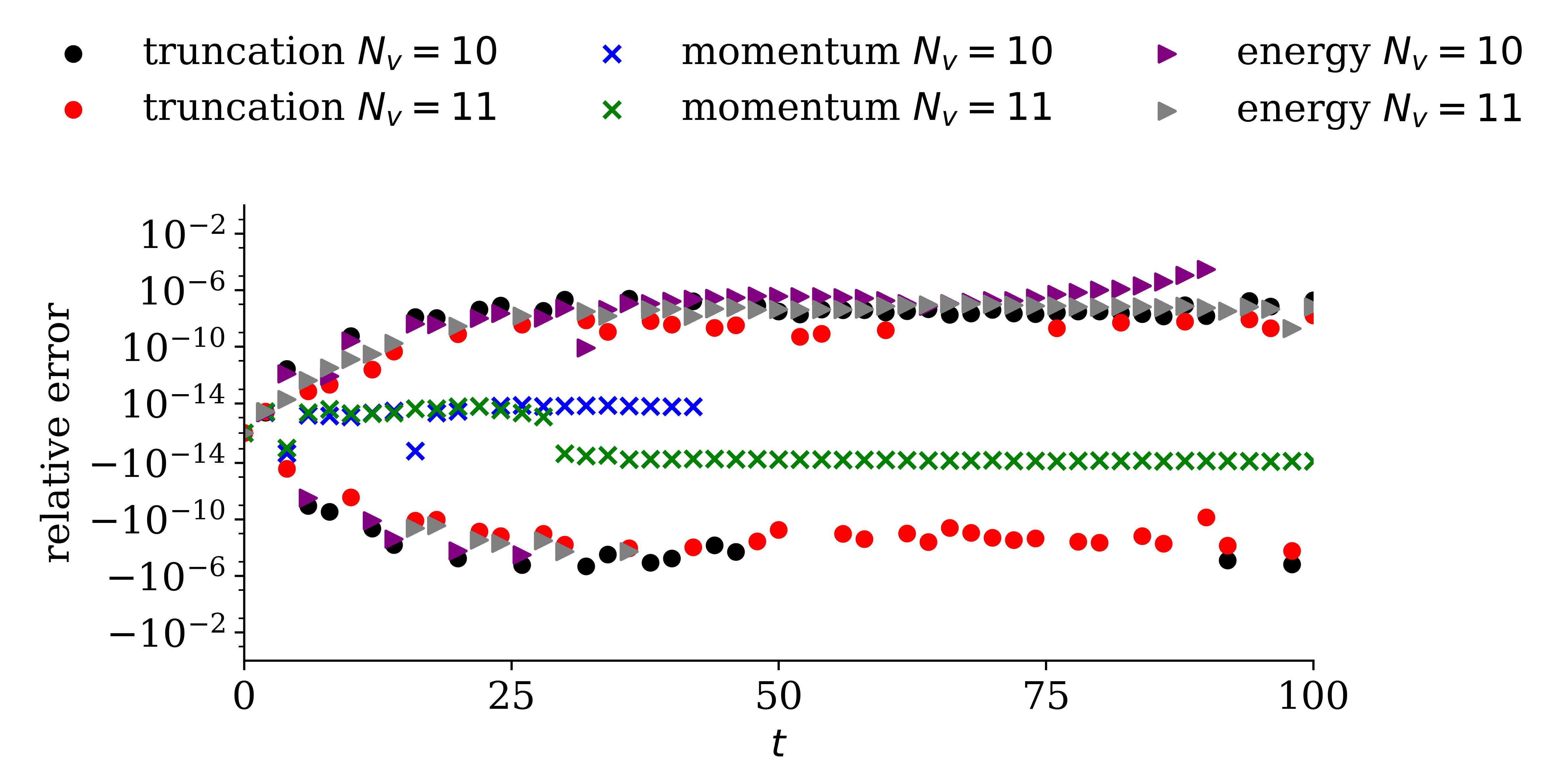}
         \label{fig:closure-by-momentum-10}
     \end{subfigure}
    \begin{subfigure}[b]{0.49\textwidth}
         \centering
         \caption{energy conservation}
         \includegraphics[width=\textwidth]{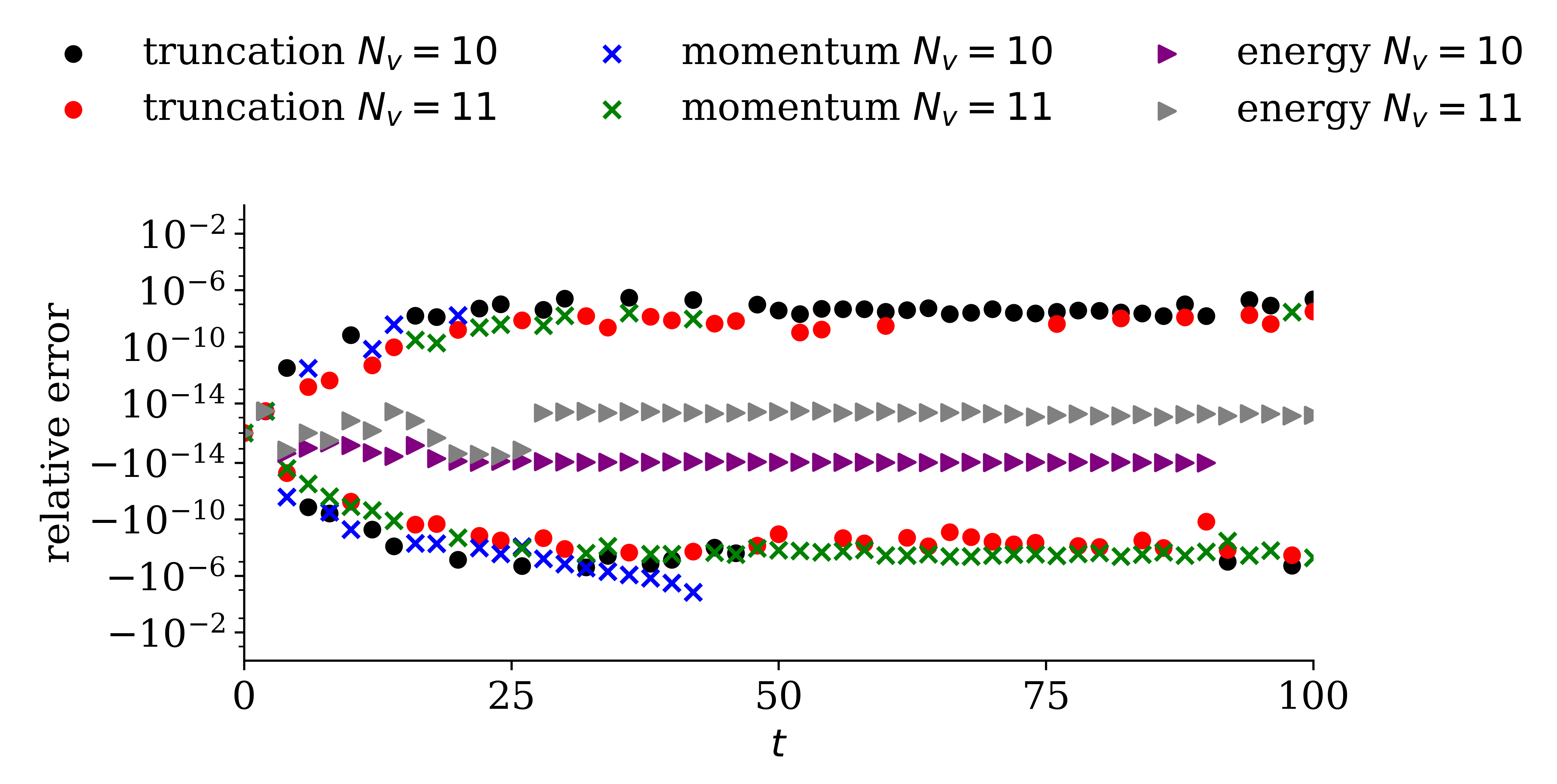}
         \label{fig:closure-by-momentum-11}
     \end{subfigure}
    \caption{Conservation relative error of total (a) $\mathcal{L}_{2}$ norm, (b) mass, (c) momentum, and (d) energy for the Langmuir wave test. The numerical results verify the analytic derivations in Eqns.~\eqref{change-in-mass-sw-closure},~\eqref{change-in-momentum-sw-closure}, and~\eqref{change-in-total-energy-sw-closure}.}
    \label{fig:conservation-langmuir-closure}
\end{figure}

%%%%%%%%%%%%%%%%%%%%%%%%%%%%%%%%%%%%%%%
\textit{How does the choice of closure influence the conservation properties?}
%%%%%%%%%%%%%%%%%%%%%%%%%%%%%%%%%%%%%%%
Figure~\ref{fig:conservation-langmuir-closure} shows the total mass, momentum, energy, and $\mathcal{L}_{2}$ norm conservation results with $N_{v}=\{10, 11\}$ for the three proposed closures: (1) closure by truncation, (2) closure by momentum, and (3) closure by energy. The numerical results agree with the analytical derivations in section~\ref{sec:closure-conservative}. Closure by truncation can conserve the $\mathcal{L}_{2}$ norm of the particle distribution function and total mass if $N_{v}$ is odd, closure by momentum can only conserve total momentum, and closure by energy can only conserve total energy. Figure~\ref{fig:conservation-langmuir-closure} indicates that the magnitude of the drift rate is comparable for the three closures. Additionally, the conservation relative error fluctuates around zero for mass, momentum, and energy, and is mainly positive for the $\mathcal{L}_{2}$ norm. The closure by momentum and energy with $N_{v}=10$ becomes unstable before the final timestamp. Recall that numerical stability properties are guaranteed only for the closure by truncation since it preserves anti-symmetry. 

\begin{figure}
    \centering
    \begin{subfigure}[b]{0.245\textwidth}
         \centering
         \caption{closure by truncation\\ \qquad w/ $N_{v}=1001$}
         \includegraphics[width=\textwidth]{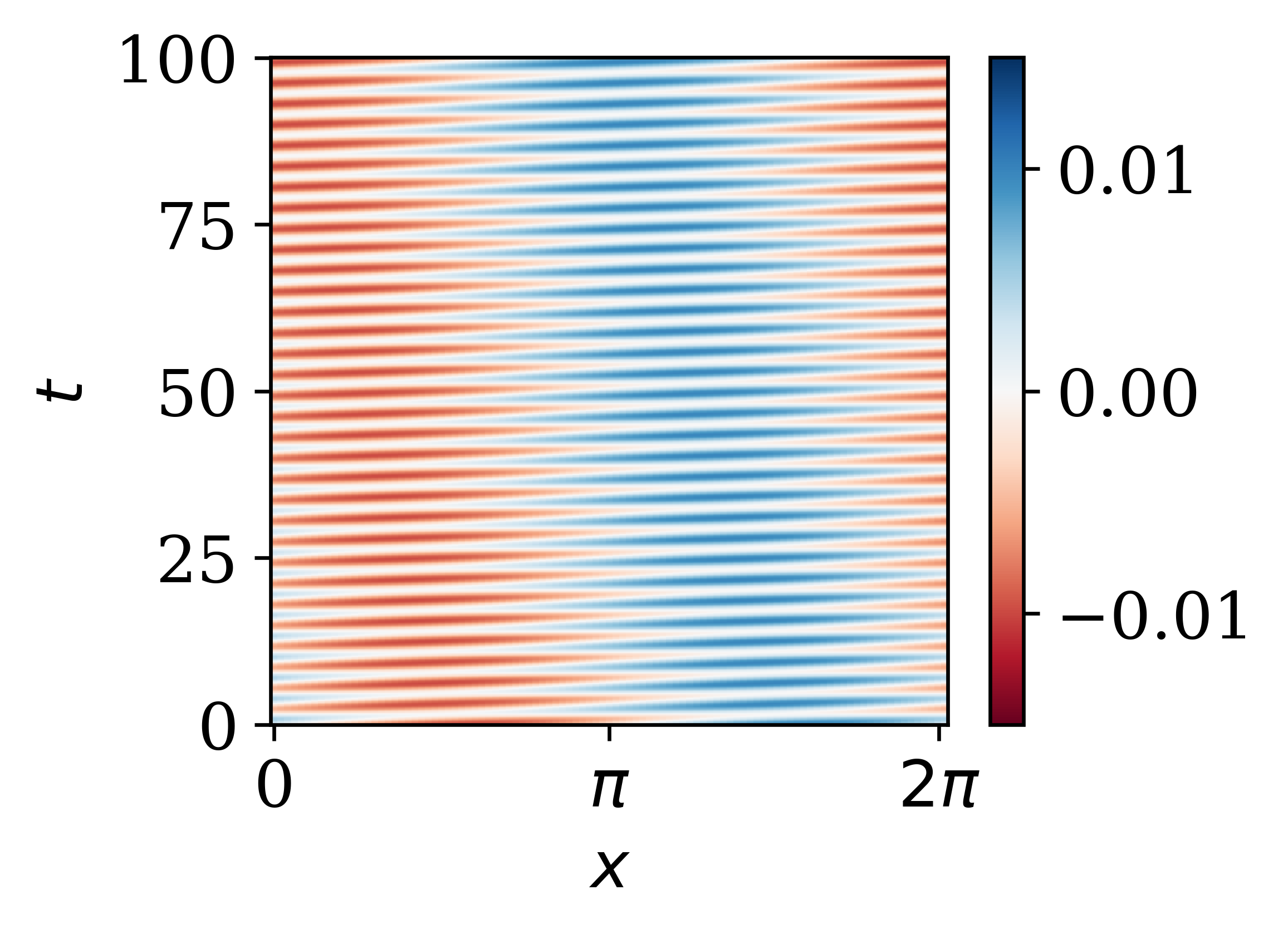}
     \end{subfigure}
        \begin{subfigure}[b]{0.245\textwidth}
         \centering
         \caption{closure by truncation\\ \qquad w/ $N_{v}=11$}
         \includegraphics[width=\textwidth]{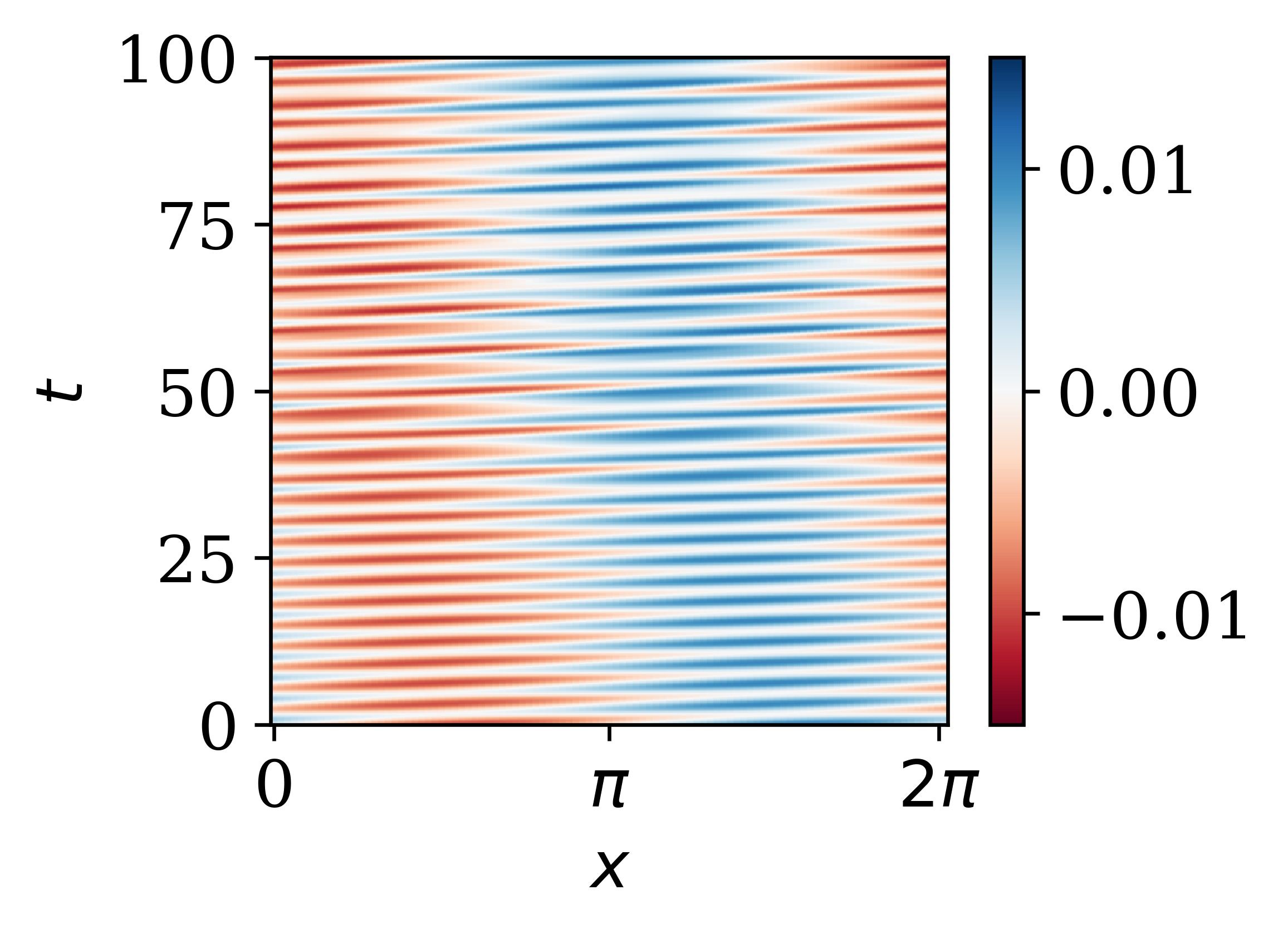}
     \end{subfigure}
    \begin{subfigure}[b]{0.245\textwidth}
         \centering
         \caption{closure by momentum\\ \qquad w/ $N_{v}=11$}
         \includegraphics[width=\textwidth]{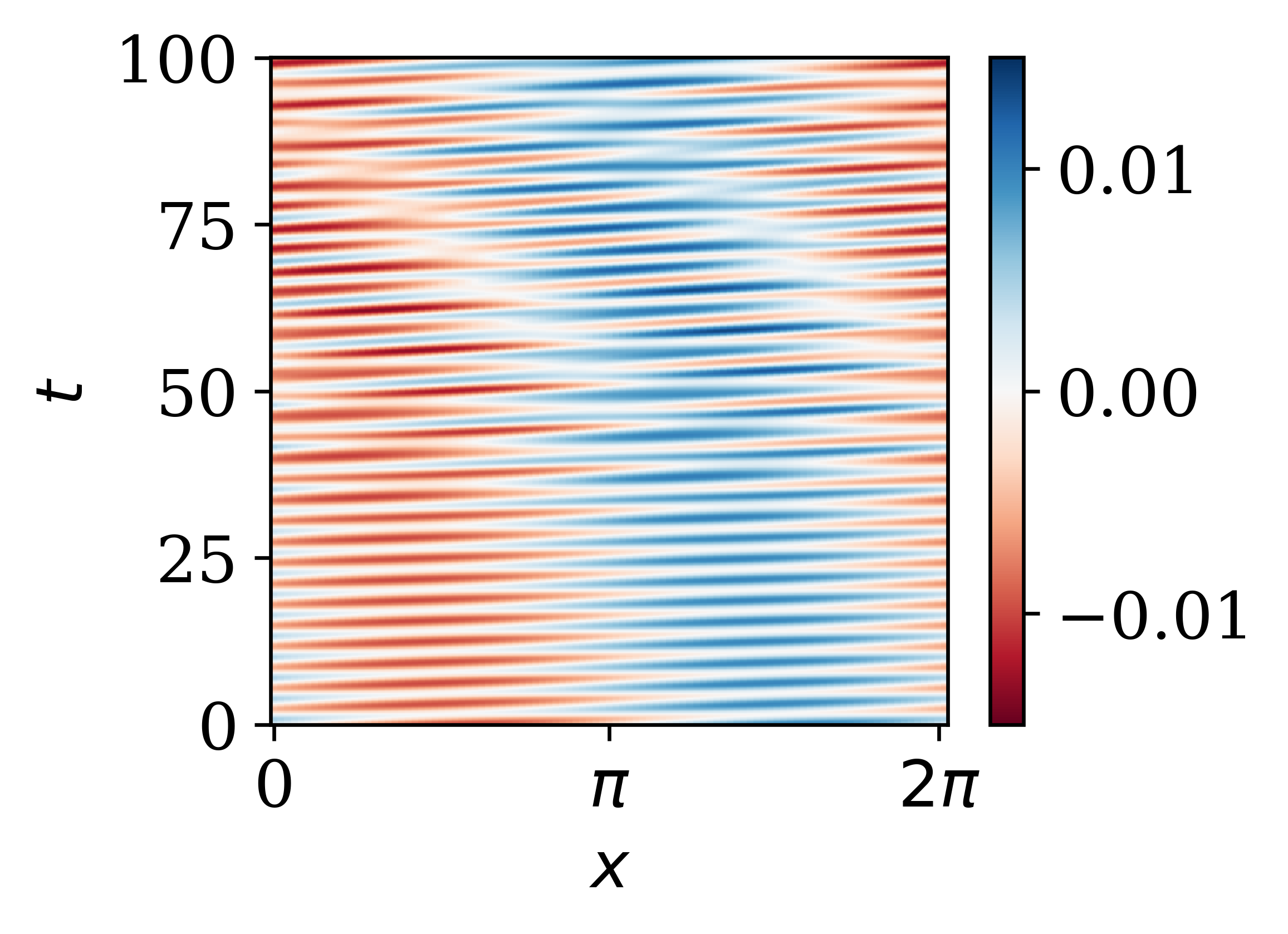}
     \end{subfigure}
    \begin{subfigure}[b]{0.245\textwidth}
         \centering
         \caption{closure by energy\\ \qquad w/ $N_{v}=11$}
         \includegraphics[width=\textwidth]{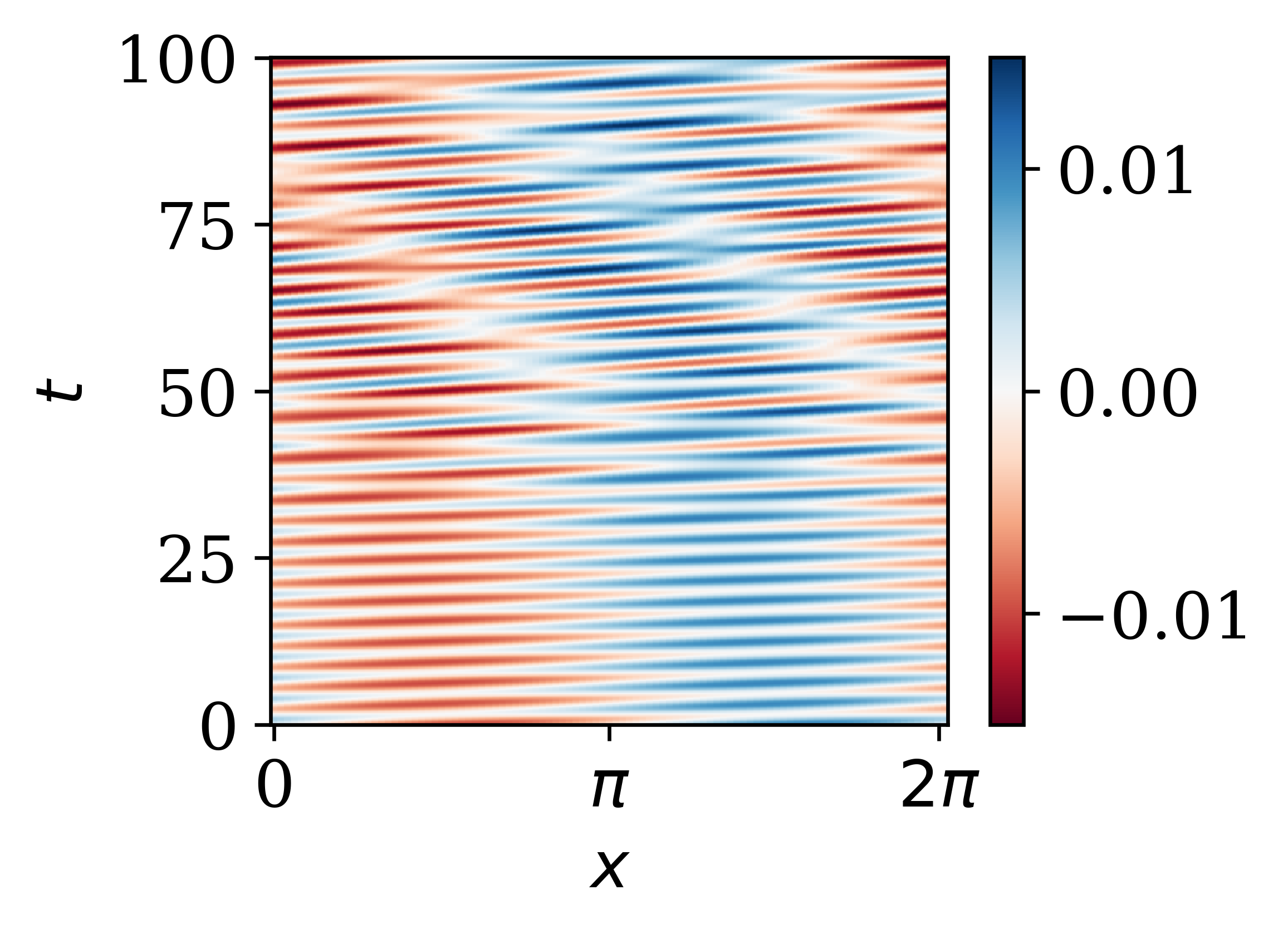}
     \end{subfigure}
     \caption{Electric field evolution comparison for the three different closures: (a) closure by truncation with $N_{v}=1001$ (reference solution), (b) closure by truncation with $N_{v}=11$, (c) closure by momentum $N_{v}=11$, and (d) closure by energy with $N_{v}=11$. The (b) closure by momentum and (d) closure by energy exhibit higher wavenumber dynamics that do not appear in the reference solution in (a). This highlights that the closure by truncation is the most suitable conservative closure for the SW Hermite method. }
     \label{fig:electric-field-comparison}
\end{figure}

%%%%%%%%%%%%%%%%%%%%%%%%%%%%%%%%%%%%%%%
\textit{How does the choice of closure impact the electric field?}
%%%%%%%%%%%%%%%%%%%%%%%%%%%%%%%%%%%%%%%
Figure~\ref{fig:electric-field-comparison} compares the electric field with $N_{v}=11$ Hermite discretization for the three closures. We compare the simulation results with a high-resolution simulation using $N_{v} = 1001$ and closure by truncation. To ensure that the choice of closure does not influence the high-resolution simulation, we verify that the magnitude of the last Hermite coefficient stays close to machine precision throughout the entire simulation. The results for the closure by momentum and closure by energy with $N_{v}=11$ are polluted by high wavenumber oscillations in comparison to the closure by truncation. This suggests that the closure by truncation may be the most appropriate conservative closure for the SW Hermite method.

%%%%%%%%%%%%%%%%%%%%%%%%%%%%%%%%%%%%%%%
\section{Concluding Remarks}\label{sec:conclusions}
%%%%%%%%%%%%%%%%%%%%%%%%%%%%%%%%%%%%%%%\
We derived conservative closures of the SW Hermite spectral solver. The presented analytic derivations show that a direct SW Hermite closure that can simultaneously conserve mass, momentum, and energy does not exist. This stems from the SW Hermite method's lack of fluid-kinetic coupling. In other words, the SW Hermite method does not satisfy the fluid equations via the first three expansion coefficients, instead, the macroscopic quantities are described by a function of all expansion coefficients. The numerical results of simulating the plasma Langmuir wave show that the closure by truncation, which assumes the last moment vanishes, is the most suitable closure for the SW formulation. This is because closure by truncation (when $N_{v}$ is odd) is the only closure that satisfies the continuity equation, and therefore preserves quasi-neutrality (locally and globally in space). Additionally, closure by truncation is the only closure that can preserve the anti-symmetric structure of the advection operator in the Vlasov equation, which is important for numerical stability purposes and conservation of the $\mathcal{L}_{2}$ norm of the particle distribution function. 

%%%%%%%%%%%%%%%%%%%%%%%%%%%%%%%%%%%%%%%
\section*{Code Availability}
%%%%%%%%%%%%%%%%%%%%%%%%%%%%%%%%%%%%%%%
The public repository~\url{https://github.com/opaliss/SW-Conservative-Closure} contains a collection of Jupyter notebooks in Python~3.9 with the code used in this study.

%%%%%%%%%%%%%%%%%%%%%%%%%%%%%%%%%%%%%%%
\section*{Acknowledgement} \label{sec:acknowledgement}
%%%%%%%%%%%%%%%%%%%%%%%%%%%%%%%%%%%%%%%
O.I. was partially supported by the Los Alamos National Laboratory (LANL) Student Fellowship sponsored by the Center for Space
and Earth Science (CSES). CSES is funded by LANL's Laboratory Directed Research and Development (LDRD) program under project
number 20210528CR. 
O.I. and B.K. were partially supported by the National Science Foundation under Award 2028125 for ``SWQU: Composable Next Generation Software Framework for Space Weather Data Assimilation and Uncertainty Quantification''. 
O.K., G.L.D., and O.I. were supported by the LDRD Program of LANL under project number 20220104DR. LANL is operated by Triad National Security, LLC, for the National Nuclear Security Administration of U.S. Department of Energy (Contract No. 89233218CNA000001).
F.D.H. was supported by the U.S. Department of Energy, Office of Science, Office of Fusion Energy Sciences, Theory Program, under Award No. DE-FG02-95ER54309. 
%%%%%%%%%%%%%%%%%%%%%%%%%%%%%%%%%%%%%%%%
% BIBLIOGRAPHY
%%%%%%%%%%%%%%%%%%%%%%%%%%%%%%%%%%%%%%%
\bibliographystyle{abbrv}
\bibliography{references}

\end{document}